\documentclass{XrU2005}
\usepackage[]{graphicx}
\usepackage[dvips]{color}
\usepackage{psfig,multicol}

\title{The XMM-Newton Slew Survey: a wide-angle survey in the 0.2 -- 12 keV band}
\author[1]{M.\,J. Freyberg}
\author[2]{B. Altieri}
\author[2]{D. Bermejo}
\author[2]{M.\,P. Esquej}
\author[2]{V. Lazaro}
\author[3]{A.\,M. Read}
\author[2]{R.\ D. Saxton}
\affil[1]{Max-Planck-Institut f\"ur extraterrestrische Physik, 85748 Garching, Germany}
\affil[2]{European Space Agency (ESA), European Space Astronomy Centre, Villafranca, 28080 Madrid, Spain}
\affil[3]{Dept.\ of Physics and Astronomy, Leicester University, Leicester LE1\,7RH, U.K.}

\newcommand{\xmm}{XMM-{\em Newton}}

\begin{document}


\maketitle

\begin{abstract}
The scientific data collected during slews of the \xmm\ satellite
are used to construct a slew survey catalogue. This comprises of the
order of 4000 sources detected in the EPIC-pn $0.2-12$\,keV band 
with exposures of less than 15\,s and a sky coverage of about 6300
square degrees (source density $\sim 0.65$ per square degree). 
Below 2 keV the sensitivity limit is comparable to
the ROSAT PSPC All-Sky Survey and the \xmm\ slew survey offers long-term
variablity studies. Above 2 keV the survey will be a factor of 10
more sensitive than all previous all-sky X-ray surveys. 
The slew survey is almost
complementary to the serendipitous survey compiled from pointed
\xmm\ observations.
It is aimed to release
the first source catalogue by the end of 2005. 
Later slew observations and detections will continuously be added.
This paper discusses the \xmm\ slew survey also in
a historical context.

\end{abstract}

\section{Introduction}

The development of new space instrumentation for X-ray astronomical applications
aims towards higher collecting areas, higher spatial resolution, and higher
spectral resolution. This is related to smaller and smaller fields of view.
Observations like {\em Deep Surveys} (e.g.\ in the directions of
the Lockman Hole, the Hubble Deep Field North, etc.) -- with exposures of the
order of $10^6$\,s
until they reach the confusion limit --
can help to study the faint end of luminosity functions and thus to analyse the
most abundant sources in the Universe.

{\em All-Sky Surveys}, on the other hand, with shallow exposures but a large sky coverage,
are the proper database to study rare objects, with a small surface number density,
and also the bright end of luminosity functions.
As an example, the ROSAT All-Sky Survey (RASS)
with its
{\em Bright Source Catalogue} with 18811 sources in the $0.1-2.4$\,keV band
(Voges et al.\ 1999a,b) exceeded any previous large-area X-ray survey
in terms of sensitivity and number of new sources.

\begin{figure}[!thb]
\centerline{\psfig{file=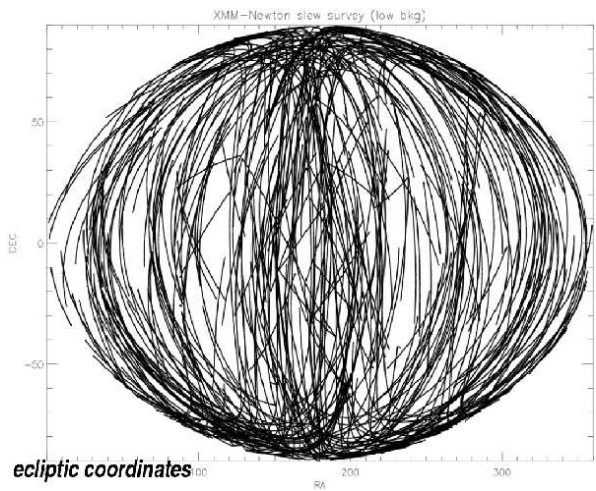,width=0.46\textwidth,angle=270.0,clip=}}%
\vspace*{-0.5mm}
\caption[]{Sky distribution of 465 EPIC-pn slews (FF, eFF, LW modes) in ecliptic
coordinates. Note that slews are performed close to great circles due to solar
angle constraints.}
\label{fig:slewpaths}
\end{figure}

{\em Slew Surveys} play an intermediate role between specially designed all-sky
programs and dedicated pointed observations.
The {\em Einstein} IPC slew survey \citep{elvis1992}
covered half of the sky in the $0.5-3.5$\,keV band with a sensitivity of
about $3\times 10^{-12}$erg\,cm$^{-2}$\,s$^{-1}$ (0.1 IPC cts/\,s$^{-1}$)
and the resulting catalogue contained 819 sources. 15\% of those had no
counterpart in the (slightly softer) RASS \citep{schachter1993}.
All-sky survey extensions into (and beyond) this harder energy range
have been proposed like ABRIXAS as a path-finder for \xmm\ \citep{truemper1998}
and ROSITA \citep{predehl2003}, but the first failed and the latter is
not yet approved. 

\xmm\ with its superior collecting area would also be an ideal mission
for serendipitous science as already in short exposures during slews
enough photons could be detected for a classification of the sources
(X-ray colours, extents, etc.). 5 years before the launch
the potential of such a slew survey was outlined \citep{lumb1995}.
Pre-launch predictions and feasibility studies, however, were based
on assumptions on the slew rate of smaller than $20^\circ$ per hour
\citep{jones1998, lumb1998, joneslumb1998}; the actual slew rate of
$90^\circ$ per hour reduces the typical number of photons per source
but increases also the sky coverage (faster slew gives more possible
observations and thus more slews).

\section{Observing strategy}

The scientific payload of the \xmm\ satellite \citep{jansen2001}
consists of
three highly nested Wolter type-I X-ray telescopes \citep{aschenbach2000}
and an Optical Monitor sensitive in the optical and UV to allow simultaneous
observations in a broad energy band up to about 12 keV.
The corresponding X-ray instrumentation is made up of 
the Reflection Grating Spectrometer (RGS).
which shares two of the three telescopes with EPIC-MOS detectors
\citep{turner2001} while behind the third telescope the EPIC-pn
camera \citep{strueder2001} receives the full intensity.
In the context of the \xmm\ Slew Survey only the imaging EPIC camera
is relevant.

\begin{figure}[!thb]
\vbox{%
\hspace*{1mm}\psfig{file=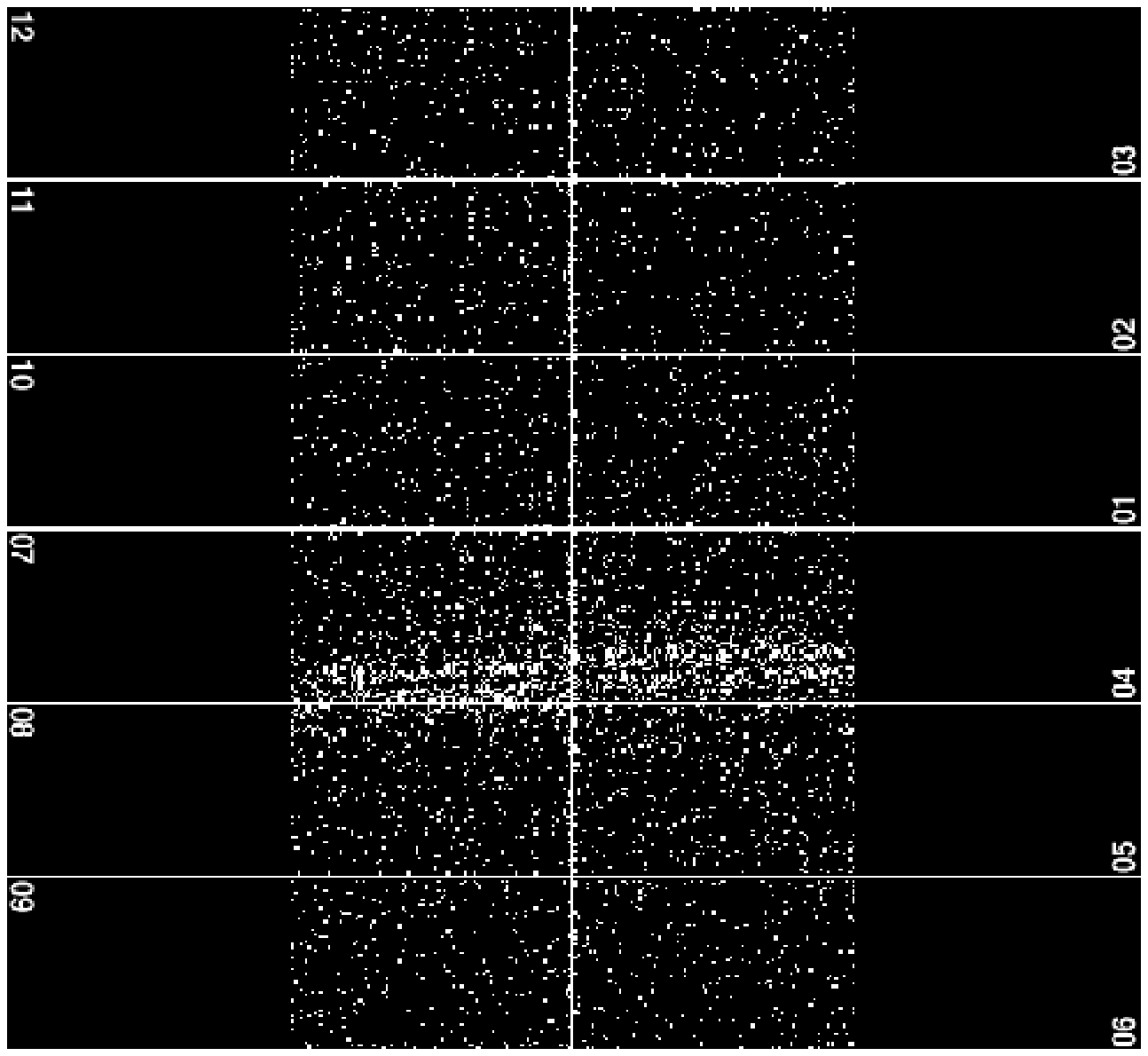,%
width=0.47\textwidth,angle=270.0,%
clip=}%
\vspace*{2mm}

\psfig{file=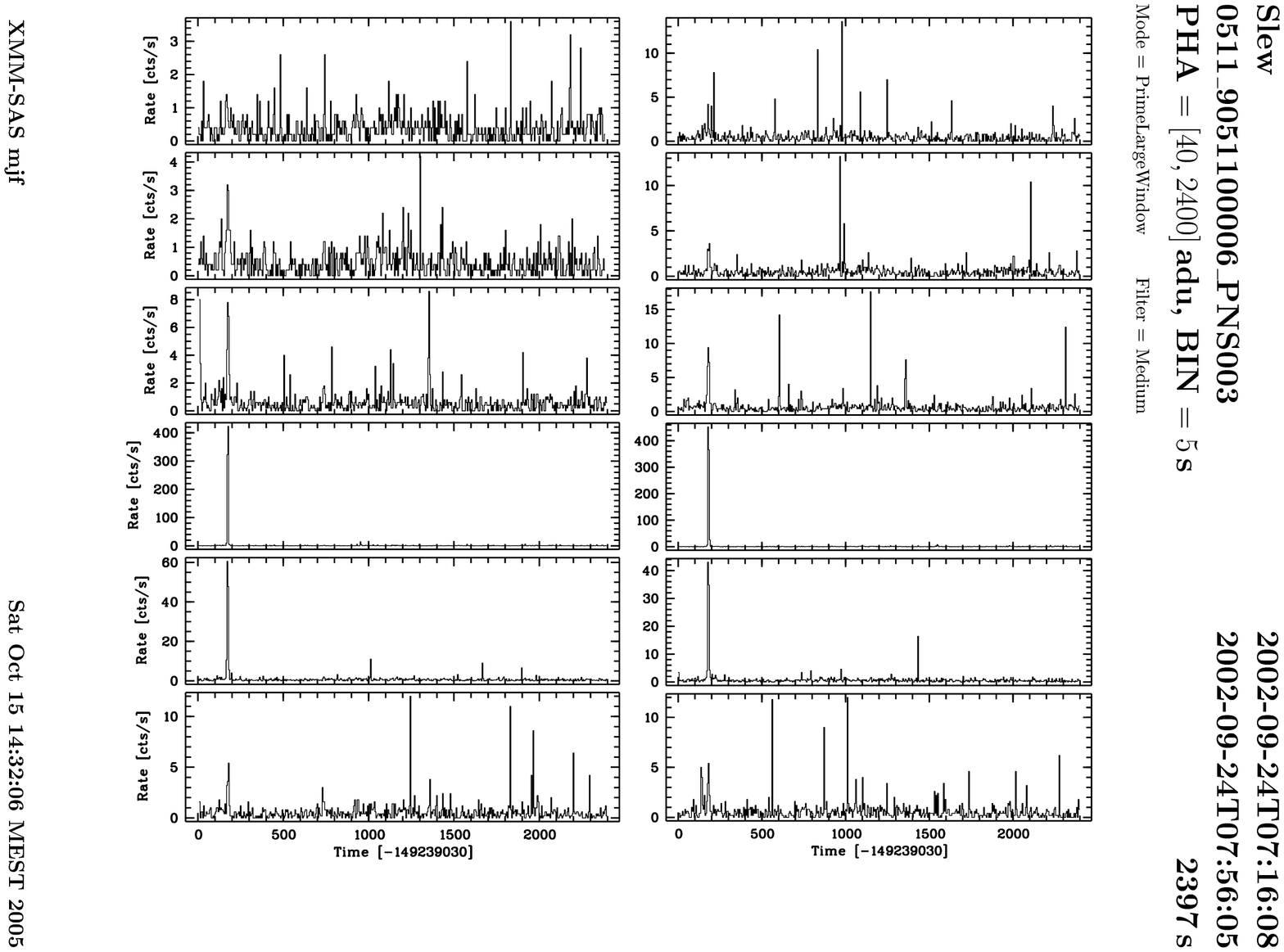,%
width=0.47\textwidth,angle=270.0,%
bbllx=61pt,bblly=149pt,bburx=513pt,bbury=678pt,clip=}%
\vspace*{2mm}

\psfig{file=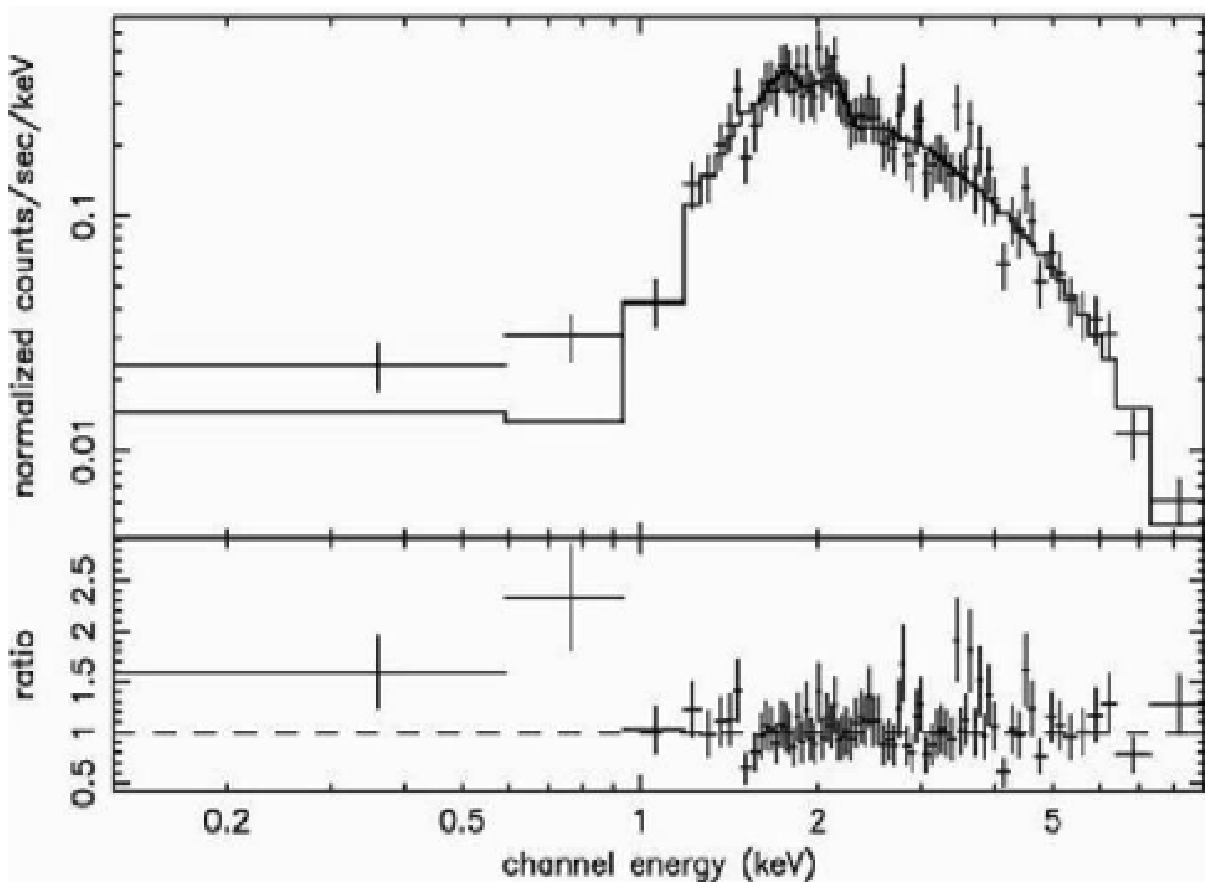,width=0.47\textwidth,angle=270,%
clip=}%
}
\caption[]{Example of a bright source found in ``pilot-0'' study. Top:
detector image, {\rm \,LW} mode. Middle: light curves of each CCD in
same orientation as above (5\,s time bins, $0.2-12$\,keV); 
the source can clearly be identified as strong
increase in CCDs 7 and 4. Bottom: source spectrum.}
\label{fig:monster}
\end{figure}

The \xmm\ mission planning tries to reduce overheads such as long slews
from one pointed observation to another. 
These slew manoeuvres between two observations and before and after
perigee passages are executed with the help of reaction wheels.
In the early phase of the mission during slews
the instruments were put into an {\tt IDLE} setup, i.e.\ they were not
completely switched off but did not collect data (except for a few
exposures with calibration filter set-up).
From revolution 314 (26 August 2001) onwards for slews lasting an hour or
longer the EPIC instruments were set into {\tt OBSERVATION} mode
with the same observation submode as the last exposure before,
and the filter wheel moved into
{\em Medium} filter position.
In particular, for EPIC-pn no new offset maps were computed
and no changes to the
uploaded bad pixel maps were applied.
Figure \ref{fig:slewpaths} illustrates the slew paths in ecliptic coordinates.
Due to solar angle constraints slews are performed close to great circles.

\begin{figure*}[!thb]
\hbox{%
\psfig{file=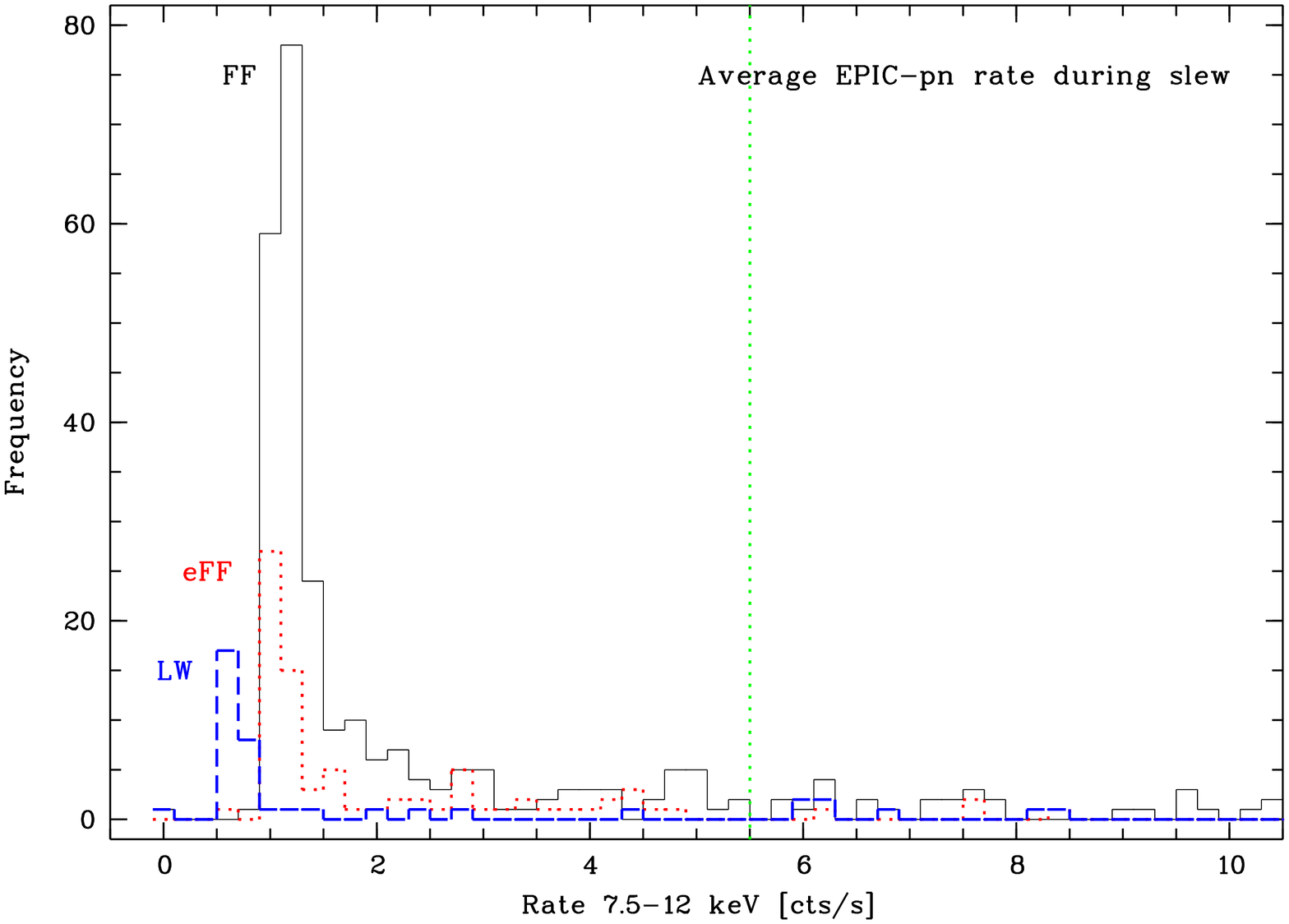,width=0.49\textwidth,angle=270.0,clip=}
\psfig{file=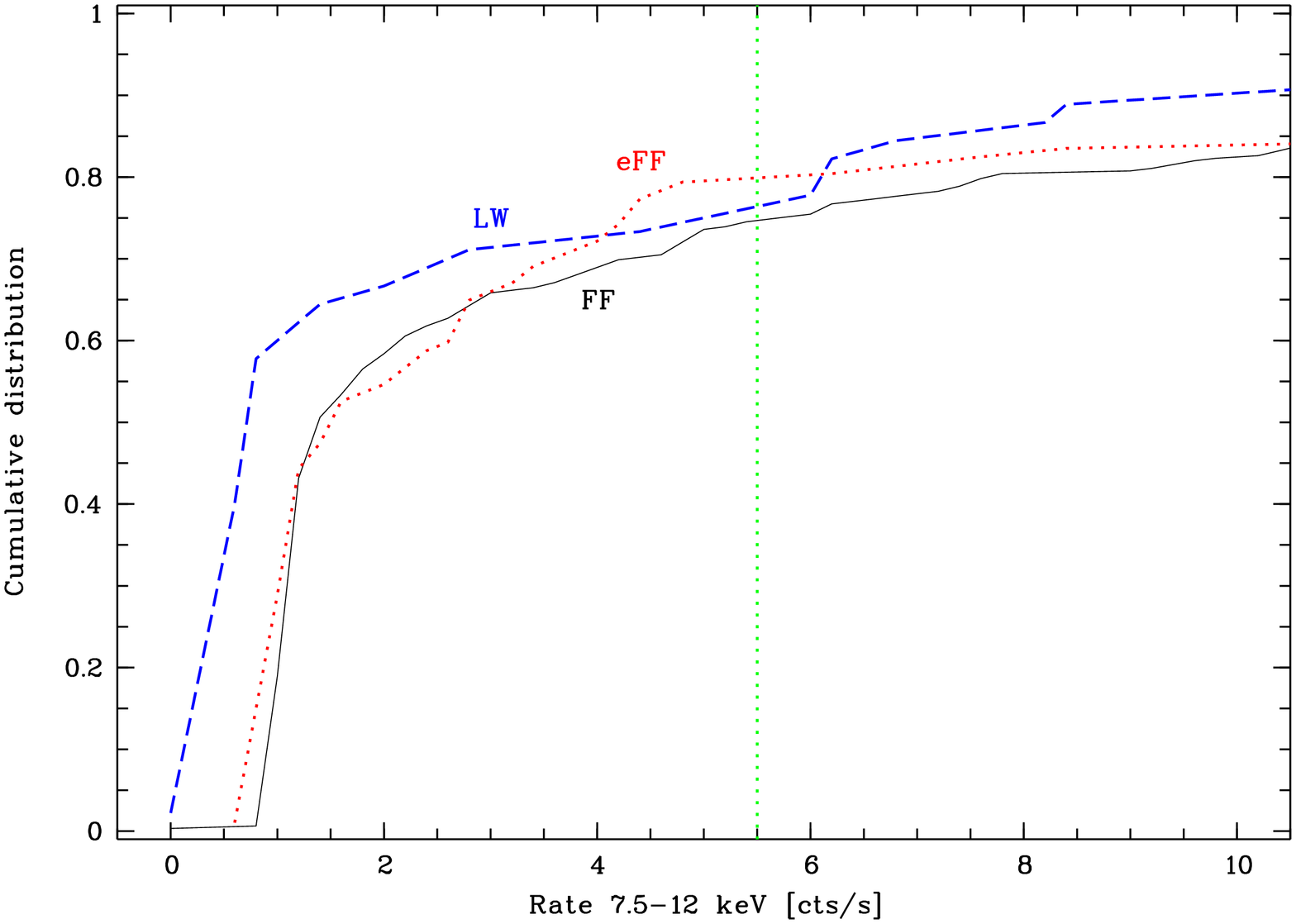,width=0.49\textwidth,angle=270.0,clip=}
}
\caption[]{Distribution of EPIC-pn count rates in the $7.5-12$\,keV band,
for the total FOV averaged over a complete slew.
Left: differential distribution for  
{\rm \,FF} (solid), {\rm \,eFF}  (dotted), and {\rm \,LW} (dashed) 
modes. Right: the corresponding cumulative
distributions. The vertical %
line indicates the threshold of
5.5 counts\,s$^{-1}$ as our selection of ``low-background slews'',
discarding about 25\% of the slews as ``high-background''.
Note, that the {\rm \,LW} mode has only about half the 
FOV of {\rm \,FF} and {\rm \,eFF} modes.}
\label{fig:rate_av_6}
\end{figure*}

The slew rate $\rho$ in the open slew phase is about $90^\circ$ per hour.
As the time resolution and thus the attitude reconstruction of CCD events 
is limited by the frame time $t_{\rm ft}$,
any image of an X-ray source scanned during a slew is distorted along
slew direction by $\rho \times t_{\rm ft}$.

\begin{table}[htb]
  \begin{center}
    \caption{``Large-scale'' EPIC imaging observation modes,
    frame times $t_{\rm ft}$, and distortion of the point spread function 
    assuming a slew rate of $\rho = 90$\,arcsec\,s$^{-1}$. For details see text.}\vspace{1em}
    \begin{tabular}[h]{ccc}
      \hline
      Mode & Frame time & Slew distortion \\
        & [s] &  [arcsec]\\\hline
      MOS FF  & 2.6 & 234 \\
      pn FF  & 0.0734 & 6.6 \\
      pn eFF  & 0.1992 & 17.9 \\
      pn LW  & 0.0477 & 4.3 \\
      \hline
      \end{tabular}
    \label{tab:ft}
  \end{center}
\end{table}

Table \ref{tab:ft} lists the frame times for all EPIC imaging modes 
where all CCDs
(7 for EPIC-MOS and 12 for EPIC-pn) are operational and in the
same mode (note, that for EPIC-MOS the outer 6 CCDs are always operated
in {\em Full Frame} mode). From this compilation it is clear that
the EPIC-pn {\em Full Frame} (FF), {\em Extended Full Frame} (eFF), and the
{\em Large Window} (LW) modes are acceptable in terms of distortion of the
point spread function while the EPIC-MOS FF mode distributes the source 
photons over a streak of about
4 arcmin during one readout cycle. Combined with the lower effective area
of EPIC-MOS the background per detection cell would be too high to add
significant value to the EPIC-pn slew data. Therefore since revolution 918
(12 December 2004)
EPIC-MOS is operated during slews with the filter wheel moved to CalClosed
position 
\citep[see also][]{harbarth2005}.
The EPIC-pn LW mode
is integrating only in the inner half of the CCD area, the field-of-view
(FOV)
is thus reduced by a factor of 2. The EPIC-pn {\em Small Window} (SW) mode
with a frame time of 5.67\,ms is only integrating in about 1/3 of the
target CCD, i.e.\ only about 3\% of the FOV, and was thus not used
by us in the compilation of the slew survey source catalogue (nevertheless,
bright sources can still be detected in EPIC-pn SW mode slews).

As the EPIC FOV is about 30 arcmin in diameter, the exposures
of slew sources reach values of up to 15\,s, depending on how central
the source passed through the FOV. For comparison, the ROSAT PSPC
had a 114 arcmin diameter FOV.

\begin{figure}[!thb]
\vbox{%
\psfig{file=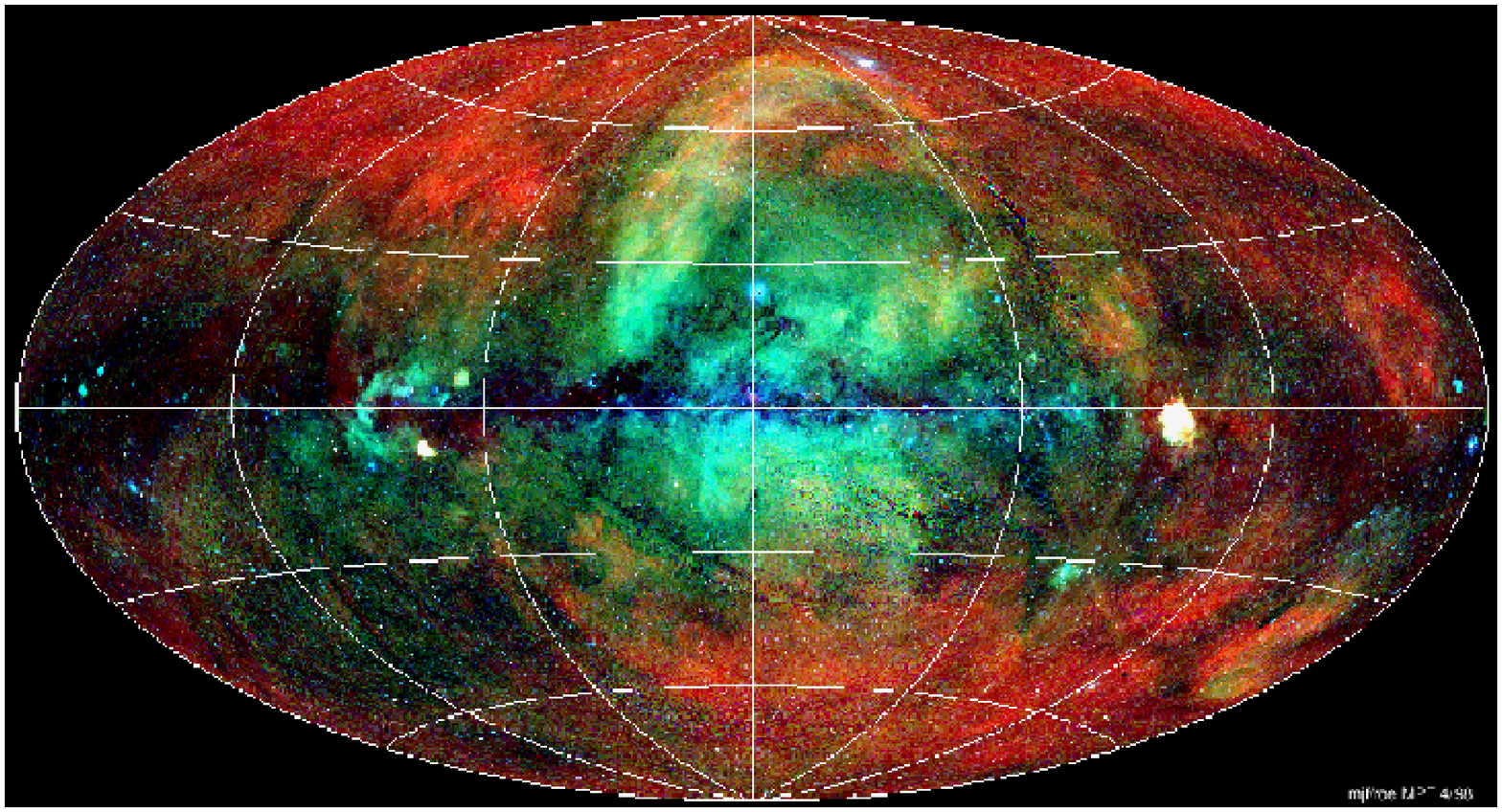,width=0.50\textwidth,angle=270.0,%
clip=}
\vspace*{2.5mm}

\psfig{file=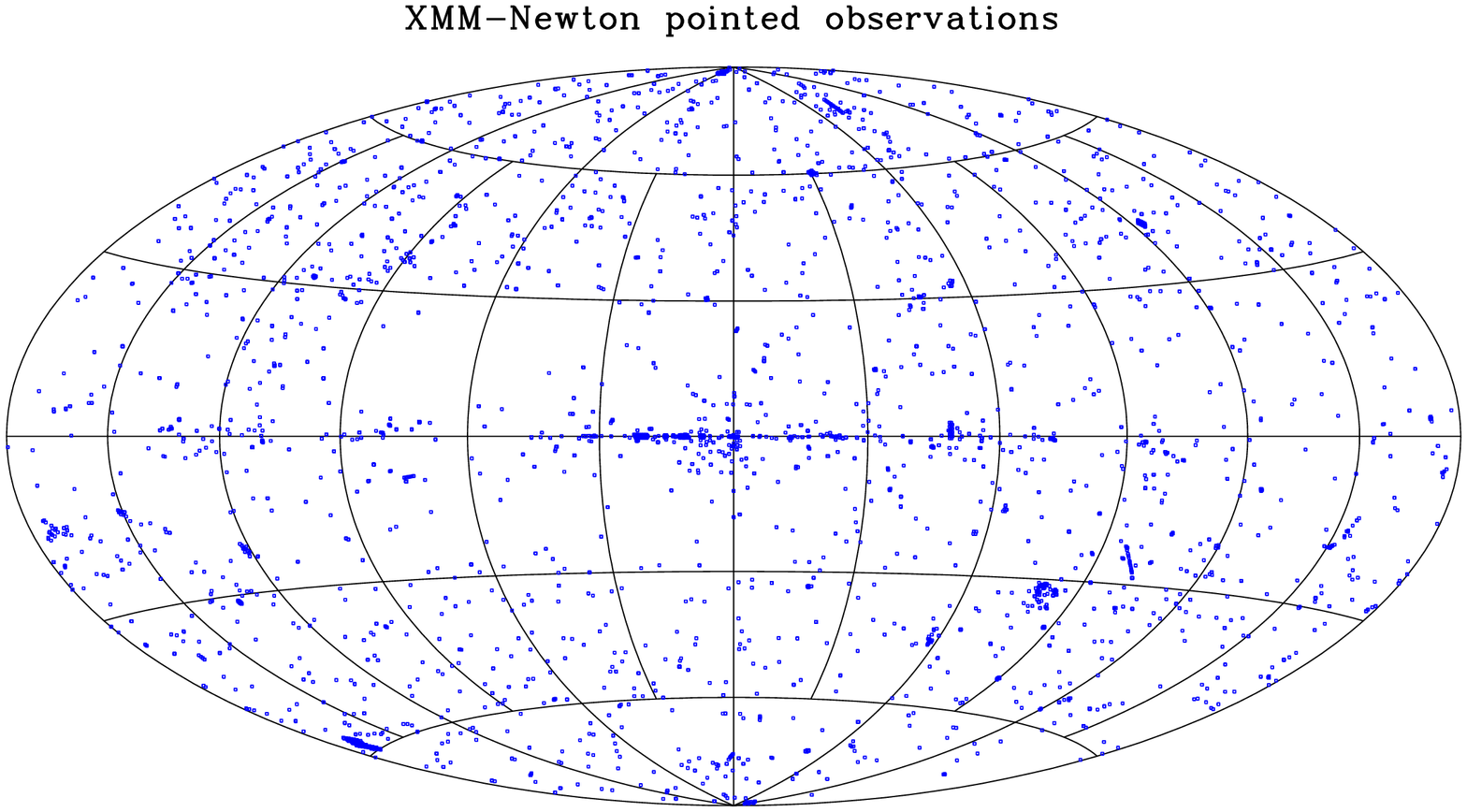,width=0.50\textwidth,angle=270.0,clip=}
\vspace*{2.5mm}

\psfig{file=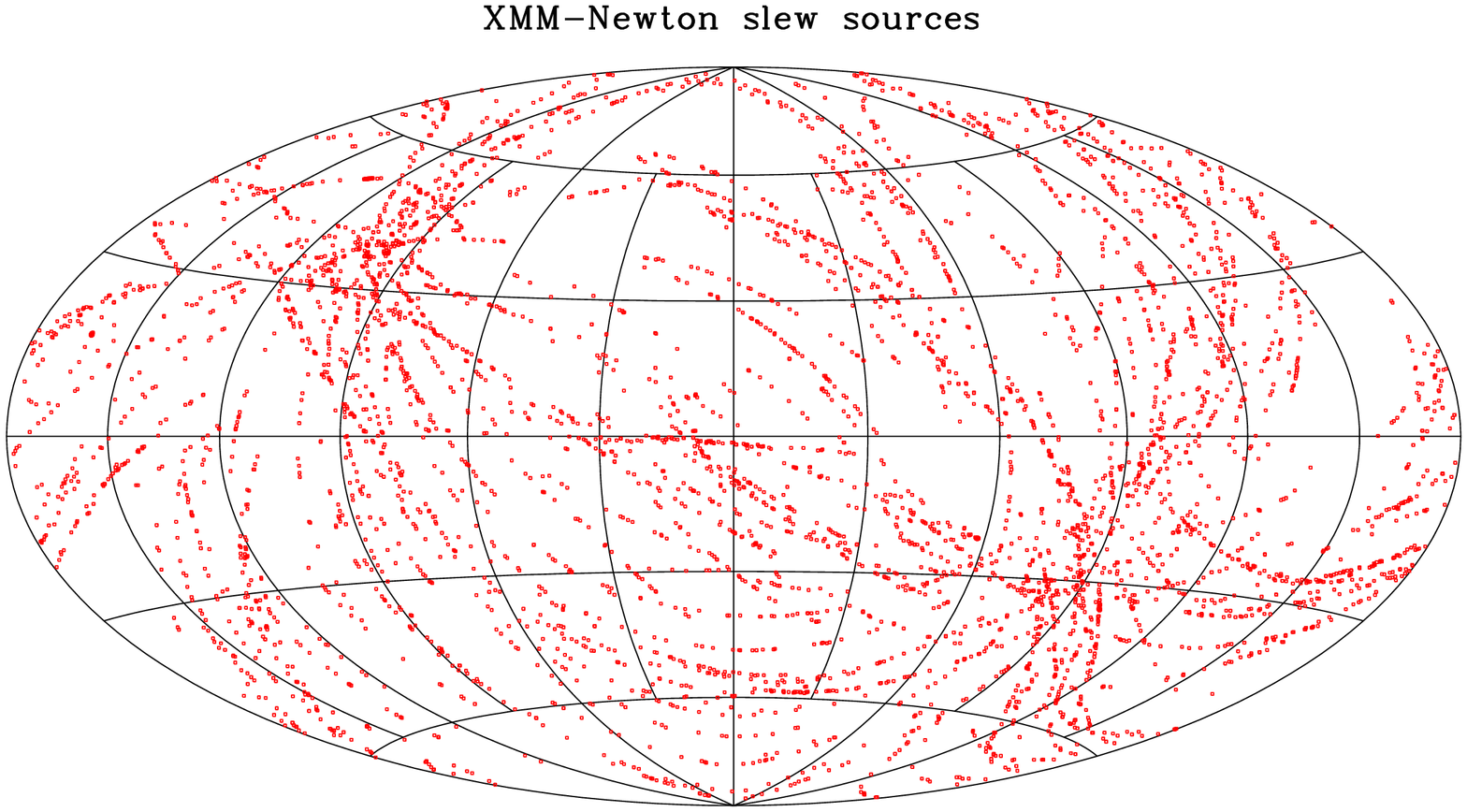,width=0.50\textwidth,angle=270.0,clip=}
\vspace*{2.5mm}

\psfig{file=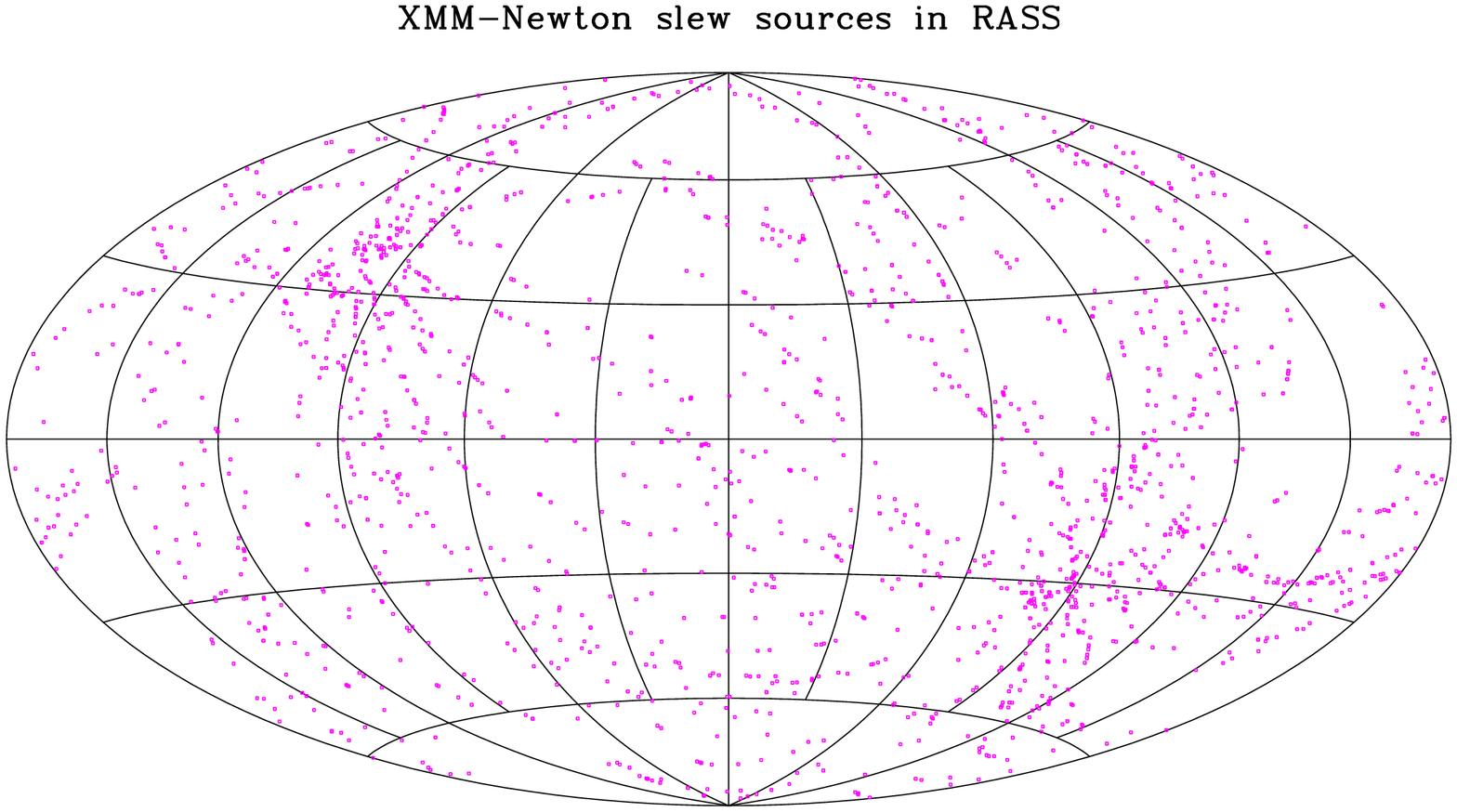,width=0.50\textwidth,angle=270.0,clip=}
}
\caption[]{Aitoff projections in galactic coordinates with galactic center in image
center and increasing longitude to the left. From top to bottom:
RASS image for comparison \protect\citep[see][and references therein]{mjfegger1999}.
Location of all XMM-Newton pointed observations.
Location of all slew survey sources (detected in the broad band); note the
different spatial distribution.
Location of all slew survey sources with a counterpart detected in the RASS.}
\label{fig:aitoff_rass_xmm}
\end{figure}

\begin{figure}[!thb]
\centerline{\psfig{file=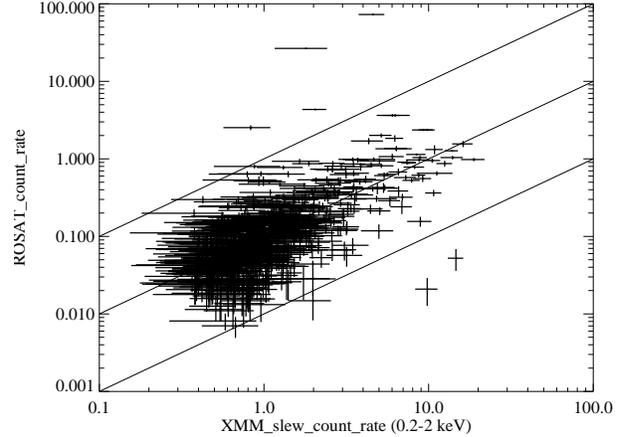,width=0.47\textwidth,angle=90.0,%
bbllx=116pt,bblly=137pt,bburx=511pt,bbury=680pt,clip=}}
\caption[]{Comparison of RASS ($0.1-2.4$\,keV) and XMM-Newton ($0.2-2.0$\,keV)
count rates for slew sources with RASS counterpart}
\label{fig:rass_xmmsl1}
\end{figure}

\section{Processing strategy}

Similarly to the {\em Observation Data Files} (ODFs) for pointed
observations there have been created
{\em Slew Data Files} (SDFs) 
with the same files and structure 
for slews not completely performed in {\tt IDLE} set-up.
These SDFs have been ingested into the {\em XMM-Newton Science Archive} (XSA).
The current public SAS ({\tt xmmsas-6.5.0}) is able to deal also correctly
with SDFs
(i.e.\ event file creation, attitude determination, exposure map computation,
source detection).

\subsection{Pilot-0 study}\label{sect:pilotzero}

As already mentioned above \xmm\ slew observations started only in August 2001.
The collected data were not immediately intended for scientific use but rather for
background studies and blank sky analysis.
In early 2004 the scientific value beyond a pure calibration aspect was
queried. In a very first pilot study (``pilot-0'')
all available SDFs were processed with the
standard SAS. Besides attitude related issues no problems occurred.

Special diagnostic products beyond the standard pipeline processing 
like detector maps in several bands and light curves of individual CCDs
were constructed. A robust source search was performed on the basis of these
light curves; a source was identified by its characteristic sequential variation 
of the rate in a number of CCDs. 
The attitude and thus the sky position was added later using
the time information. It was shown that in all EPIC-pn imaging modes
(FF, eFF, LW, and SW) sources can be detected and related to other
catalogued objects.

For all SDFs background lightcurves and average rates in 6 bands
for the total FOV were computed.
The distribution of these rates for the highest band chosen
($7.5-12$\,keV) is shown in Fig.\ref{fig:rate_av_6}. The left panel contains the
differential distribution for 
FF (solid), eFF (dotted), and LW (dashed) 
modes, the right panel the corresponding cumulative
distributions. The vertical %
line indicates the threshold of
5.5 counts\,s$^{-1}$ as our selection of ``low-background slews'',
discarding about 25\% of the slews as ``high-background''.
Note, that the LW mode has only about half the FOV of the FF and eFF modes
and that therefore the count rates are lower by a factor of about 2.

\subsection{Pilot-1 study}\label{sect:pilotone}

In a following study (``pilot-1'') a number of slews in FF mode were analysed
to verify effects of optical loading and to determine the best settings
for image creation, energy bands, and source detection via standard SAS tasks
\citep{read2005, saxton2005}.
As the tangential plane projection is not valid anymore over the whole
slew, it was split into event files of 1 square degree size and
resulting sub-images were used for source detection.
Sub-images with exposures of $> 35$\,s (i.e.\ close to the end of the slew)
were discarded from the pipeline.
It was concluded to use slew data in FF, eFF, and LW
modes, and to drop SW mode exposures from the slew survey catalogue processing.

\subsection{Pilot-2 study}\label{sect:pilottwo}

Three long low-background slews (one in FF, eFF, and LW modes, close-by in time)
were specifically
analysed to identify any mode-dependent features in the source detection
pipeline and whether a common scheme could be applied to all EPIC-pn slew data.
Aspects of this pipeline have been described in detail by
\citet{read2005} and \citet{saxton2005}.

\subsection{Current scheme}\label{sect:current}

From the abovementioned pilot studies the following processing scheme
for SDFs was set up \citep[see also][]{esquej2005}:\vspace*{-2mm}
\begin{itemize}
\item produce event files down to 100\,eV (to be able to later identify
optical loading, detector flashes etc.)
\item select only FF, eFF, and LW mode slew exposures
\item discard slews with average rate in $7.5-12$\,keV band exceeding
  5.5\,counts\,s$^{-1}$
\item construct images and exposure maps in 3 bands: soft (``RASS'') band
$0.2-2.0$\,keV, hard (``beyond RASS'') band $2-12$\,keV, total band $0.2-12$\,keV.
Note, that in the range $0.2-0.5$\,keV only single pixel events are used 
({\tt PATTERN==0}) and above 0.5\,keV also double pixel events ({\tt PATTERN.le.4}),
and discard subimages with exposures $>35$\,s.
\end{itemize}

\section{The Catalogue: XMMSL1}

The \xmm\ Slew Survey catalogue XMMSL1 will consist of the order of 4000 sources
detected in the total band ($0.2-12$\,keV), about 2700 in the soft
and about 800 in the hard bands alone, respectively.
After flagging and removal of spurious sources 
(e.g., due to detector artifacts, optical loading, software) --
which will reduce the figures mentioned above --
the catalogue is planned to be released by the end of 2005 \citep{esquej2005}.

\subsection{Quality control}

Quality control is a very important issue to ensure valuable scientific exploitation,
and started already with the event file creation. The {\tt verbosity=5}
log file of the task {\tt epchain} was inspected for messages indicating
possible unusual features in detector performance or data flow.
The special data products described in Sect.\ref{sect:pilotzero}
were very useful for sanity checks.
Searches for optical loading and detector artifact cases
were performed below the actual slew survey limit (0.2 keV).

An extensive cross-correlation of \xmm\ slew sources with various
other X-ray and optical catalogues was performed: faint detections without such
counterparts
have a higher probability of being spurious. 
Figure \ref{fig:aitoff_rass_xmm} shows the distribution of \xmm\ 
pointed observations, of slew sources, and of slew sources with a detection
in the RASS.
In Fig.\ref{fig:rass_xmmsl1} we show the relation of soft band slew source rates
with RASS rates, where a general correlation is observed (``XMM $= 10\times$ RASS'')
with scatter due to different instrument responses in the energy bands, and also due to
long-term variability, but may also point towards residual uncertainties due
to optical loading or event pattern pile-up.
We have also checked the slew survey sources that have {\em no} counterpart
in the RASS and have determined upper limits for the detection to identify
further candidates for visual inspection of spurious sources.

The slew survey is almost
complementary to serendipitous surveys compiled from pointed
\xmm\ observations \citep[see, e.g.,][for the AXIS programme]{barcons2002}. 
This is mainly due to a selection bias because the
sky portion observed in the pointing at the start and at the end of a slew
is not part of the slew survey. It did, however, happen that a slew
passes over a field that is part of the pointed programme at a different phase.
These cases can also be used for variability studies.

\begin{figure}[!thb]
\centerline{\psfig{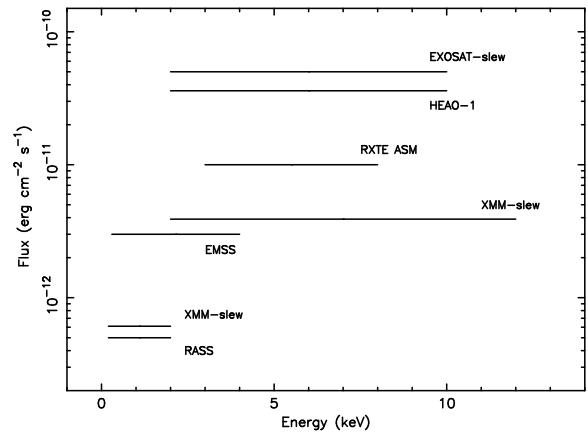}}
\caption[]{Sensitivity limits for various X-ray surveys.}
\label{fig:sensitivities}
\end{figure}

\subsection{Survey sensitivities}

The short exposure times in the \xmm\ slew survey are partly
compensated by the superior collecting area of the X-ray telescopes.
Below 2 keV the sensitivity limit 
of $6 (4.5) 10^{-13}$\,erg\,s$^{-1}$\,cm$^{-2}$
(for detection likelihoods of 10 and 8, respectively)
is comparable to
the ROSAT PSPC All-Sky Survey, and the \xmm\ slew survey offers long-term
variablity studies. Above 2 keV the survey will be a factor of 
2.5
more sensitive 
($4 (3) 10^{-12}$\,erg\,s$^{-1}$\,cm$^{-2}$)
than the RXTE survey \citep{revnivtsev2004} -- which has a positional
accuracy of only $\sim~1^\circ$ --
and a factor of 10 compared 
to other (spatially resolved) large-area X-ray surveys (Exosat, HEAO-1).
Figure \ref{fig:sensitivities}
compares sensitivity limits of previous surveys with the current
\xmm\ slew survey.

The RASS showed large exposure inhomogeneities due to the survey design:
all slews scanned over the ecliptic poles leading to a significantly
higher exposure there than in the ecliptic plane. The \xmm\ slew survey
is not that strongly biased to the ecliptic poles: Fig.\ref{fig:slewpaths}
shows instead a wide scatter of slew paths, and due to the significantly
smaller FOV (factor $\sim 14$) only a number of crossings of two slew paths
and very occasionally of three slew paths (there are few slew point sources that have
been detected in three different slews).

When constructing luminosity functions from slew survey catalogues one has
to keep in mind selection effects. As slew paths start and end just next
to targets of pointed observations, these preferentially brighter sources
are included in the slew survey with lower probability: there is a small
bias against bright sources. Including the target area would turn the
bias into an overabundance of brighter sources.

Unlike in the RASS where each part of the sky was scanned multiple times
(survey rate in the ecliptic plane $\sim 4$\,arcmin per orbit with a FOV
of 114 arcmin diameter) the exposure strongly depends on the off-axis angle
in the \xmm\ slew. The \xmm\ survey sensitivity is therefore strongly inhomogeneous
perpendicular to the slew direction (rather than perpendicular to the ecliptic plane).

\section{Conclusions and Outlook}

It has been shown that the \xmm\ EPIC-pn data collected during slews
represent an important scientific database. The catalogue currently
under construction will provide a complement to catalogues compiled
from pointed observations (1XMM, 2XMM).
Moreover, not only point sources but also extended sources had been
detected in the slew survey \citep{lazaro2005}. While supernova remnants
are most likely already detected in the RASS, harder sources like clusters
of galaxies may be new extended objects originating from the slew survey.

In further versions of the slew survey catalogue it is planned to 
recover part of the slews that were disregarded due to high background by
selecting periods of low background (using good time intervals similar to
pointed observations).
The slew sky coverage will increase and therefore serendipitous overlaps with 
pointed observations and with other slew paths will increase as well.
This will greatly enhance the possibility of time variability studies.

\section*{Acknowledgments}

The XMM-Newton proj\-ect is an ESA Science Mission with in\-stru\-ments
and contributions directly funded by ESA Member States and the USA (NASA).
The XMM-Newton project is supported by the
Bundesministerium f\"ur Bildung und For\-schung/Deutsches Zentrum
f\"ur Luft- und Raumfahrt (BMBF/DLR),
the Max-Planck-Gesellschaft, and the
Hei\-den\-hain-Stif\-tung.

\end{document}